\documentclass[graybox]{svmult}

\usepackage{mathptmx}       
\usepackage{helvet}         
\usepackage{courier}        
\usepackage{type1cm}        
\usepackage{makeidx}         
\usepackage{graphicx}        
\usepackage{multicol}        
\usepackage[bottom]{footmisc}

\makeindex             

\begin{document}

\title*{Discussion on ``Techniques for Massive-Data Machine Learning in Astronomy" by A. Gray}
\author{Nicholas M. Ball}
\institute{Nicholas M. Ball \at National Research Council Herzberg Institute of Astrophysics, 5071 West Saanich Road, Victoria, BC V9E 2E7, Canada \email{nick.ball@nrc-cnrc.gc.ca}}
%
%
\maketitle

\section{Introduction}
\label{sec:intro}



Astronomy is increasingly encountering two fundamental truths:

\begin{itemize}
\item{The field is faced with the task of extracting useful information from extremely large, complex, and high dimensional datasets.}
\item{The techniques of {\it astroinformatics}\cite{ball:ijmpd,borne:datamining}\footnote{\url{http://www.ivoa.net/cgi-bin/twiki/bin/view/IVOA/IvoaKDDguide}} and {\it astrostatistics} are the only way to make this tractable, and bring the required level of sophistication to the analysis.}
\end{itemize}

Thus, an approach which provides these tools in a way that scales to these datasets is not just desirable, it is vital. The expertise required spans not just astronomy, but also computer science, statistics, and informatics. As a computer scientist and expert in machine learning, Alex's contribution of expertise and a large number of fast algorithms designed to scale to large datasets, is extremely welcome. We focus in this discussion on the questions raised by the practical application of these algorithms to real astronomical datasets. That is, what is needed to maximally leverage their potential to improve the science return?

This is not a trivial task. While computing and statistical expertise are required, {\it so is astronomical expertise}. Precedent has shown that, to-date, the collaborations most productive in producing astronomical science results (e.g, the Sloan Digital Sky Survey), have either involved astronomers expert in computer science and/or statistics, or astronomers involved in close, long-term collaborations with experts in those fields. This does not mean that the astronomers are giving the most important input, but simply that their input is crucial in guiding the effort in the most fruitful directions, and coping with the issues raised by real data. Thus, the tools must be useable and understandable by those whose primary expertise is not computing or statistics, even though they may have quite extensive knowledge of those fields.



`Real' astronomical data are characterized by many issues which differentiate them from ideal data. They may:

\begin{itemize}
\item{Be large, complex, increasingly high-dimensional, and may be in the time domain}
\item{Contain missing data, such as non-observations or non-detections}
\item{Have heteroscedastic (changing variance), non-Gaussian, or underestimated errors}
\item{Contain outliers, artifacts, false detections, or systematic effects}
\item{Contain correlated inputs}
\item{... and so on}
\end{itemize}

\section{Relevance of the algorithms presented}
\label{sec:relevance}

The algorithms presented meet the criteria of being well-known (kNN, KDE, etc.), scalable ($N$log$N$ where possible), and useable by astronomers via the software of the FASTlab group. Some of the well-known algorithms already scale without the work of the group, e.g., mixture of Gaussians, decision tree, linear regression, K-means, and PCA. However, others, such as all nearest neighbors, KDE, SVM, and nPCF, do not. What is significant about the results presented here is that they make all of these algorithms scalable. Extensive use is made of the fact that to build a $kd$-tree data structure scales as $N$log$N$. This and other space-partitioning tree structures are what makes the scaling possible.

The relevance of the work of the group is two-fold: (a) their results enable scalable versions of the algorithms that do not otherwise scale to be implemented; and (b) they give one the ability to employ more sophisticated variants of the algorithms that do scale. For example, many astronomical phenomena, such as galaxy spectra, are nonlinear, but are often treated by linear analyses such as PCA, or templates. Kernel PCA is a nonlinear extension of PCA, and in the results presented scales as O($N$), rather than O($N^3$). There are numerous other examples. Both of these points increase the applicability of the algorithms to real astronomical data, i.e., data that contains the issues listed in section \ref{sec:intro}.

\section{CADC, CANFAR, Petascale Data, and Fast Data Mining Algorithms}
\label{sec:cadc}

The Canadian Advanced Network for Astronomical Research (CANFAR) \cite{gaudet:canfar} is a project at the Canadian Astronomy Data Centre (CADC) to provide an infrastructure for data-intensive astronomy projects. It provides those portions of a pipeline that can be usefully supplied in a generic manner, such as access to, processing, storage, and distribution of data, without restricting the analysis that can be performed. The system combines the job scheduling abilities of a batch system with cloud computing resources, and users manage one or more virtual machines, which operate (to them) in the same manner as a desktop machine.

By extension of the arguments for providing a hardware infrastructure and standard software tools within CANFAR, we aim to provide a robust set of generic tools that can be used for data analysis. Given the requirements detailed in section \ref{sec:intro}, that the methods of astroinformatics and astrostatistics are needed for appropriately sophisticated analysis of the data, that such algorithms must scale as $N$log$N$ or better to remain tractable in the upcoming petascale regime, and that the aim of the FASTlab group is to implement them such that they may be used on real problems, we are using the software of the group to achieve our aims.

The key point is that, while a given science analysis always specific, {\it the underlying algorithms are generic}, and it is those that we aim to provide.

\section{Example: The Next Generation Virgo Cluster Survey}
\label{sec:ngvs}


The Next Generation Virgo Cluster Survey (NGVS)\footnote{\url{https://www.astrosci.ca/NGVS/The\_Next\_Generation\_Virgo\_Cluster\_Survey}} is a new 104 square degree survey of the Virgo Cluster, which will provide coverage of this nearby dense environment in the universe to unprecedented depth. The limiting magnitude of the survey is $g_{AB} = 25.7$ ($10\sigma$ point source), and the $2\sigma$ surface brightness limit is $g_{AB} \approx 29~\mathrm{mag~arcsec}^{-2}$. The data volume of the completed survey will be approximately 50 terabytes. The objects detected span an enormous dynamic range, from the giant elliptical galaxy M87 at $M(B) = -21.6$, to the faintest dwarf ellipticals at $M(B) \approx -6$. Photometry will be available in 5 broad bands ($u$* $g$' $r$' $i$' $z$'), and the unprecedented depth reveals many complex and previously unseen low surface brightness structures. Some of the survey challenges are given in Table \ref{tab:1}, together with the relevant machine learning algorithm, and the speedup provided by the results of the FASTlab group.

\begin{figure}[b]
\includegraphics[scale=0.325]{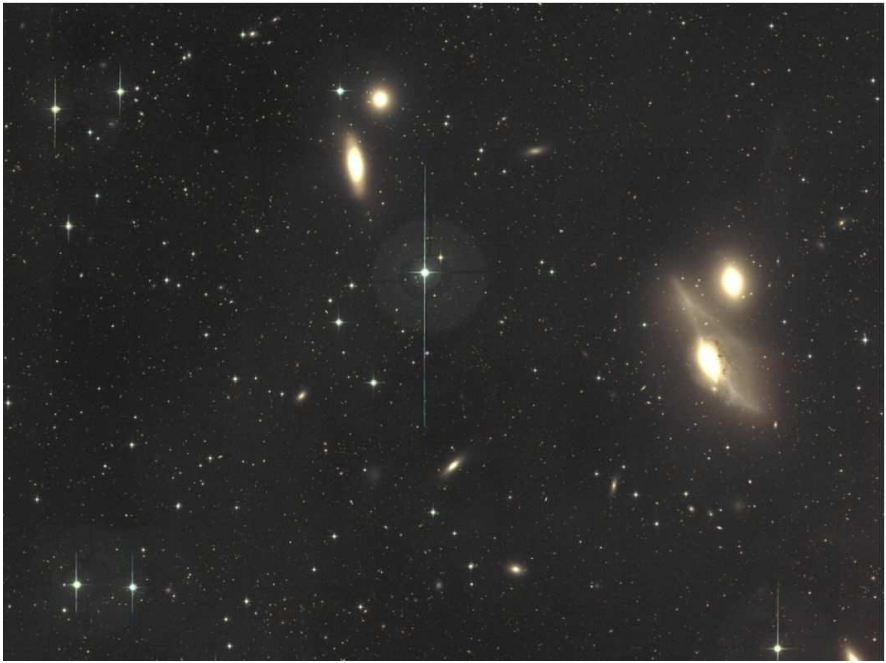}
\caption{Typical NGVS survey region, showing several challenges to data mining provided by this survey, including: (1) full-colour images, provided by 5-band photometry; (2) bright stars, exhibiting halos and bleed trails; (3) large galaxies, showing elliptical light profiles, colour gradients, and detailed morphology; (4) complex, irregular, galaxy morphologies - the galaxy on the right is NGC 4438; (5) similar low-surface brightness features, the incidence of which is hugely increased by the survey's unprecedented depth; (6) low surface brightness galaxies, e.g., below NGC 4438; (7) globular clusters and ultra-compact dwarfs - these objects may be unresolved, or partially resolved, and exhibit different light profiles to galaxies, complicating their classification, and the separation of stars (unresolved) and galaxies (resolved); (8) varying sky background, especially near large galaxies, whose light extends to large radii; (9) most objects in the image have no spectroscopy, thus their membership, or not, of the Virgo Cluster, must be deduced by other means.}
\label{fig:1}
\end{figure}

A typical region of the survey is shown in Figure \ref{fig:1}, further exemplifying some of the challenges, and adding others. Many of these, which do not make direct use of the algorithms, but rather of other astronomical software, may be sped up by a linear factor equal to the number of processing cores (currently several hundred) available on the CANFAR system.

Thus, the combination of the fast algorithms provided by Alex's group, and the CANFAR system, enables large datasets to become tractable, while at the same time, for challenges that the algorithms do not directly address, enabling those too to be tackled. Thus, the revolutionary, but nevertheless real and not idealized astronomical data of the NGVS and future surveys, is being tackled in a smart, and scalable way.


\begin{table}
\caption{NGVS tasks and FASTlab speedups (potential or actual)}
\label{tab:1}
\begin{tabular}{p{5.3cm}p{2cm}p{2cm}p{2cm}}
\hline\noalign{\smallskip}
Task & Algorithm & Naive speed & FASTlab speed \\
\noalign{\smallskip}\svhline\noalign{\smallskip}
Object classification & SVM & O($N^3$) & O($N$) \\
Virgo Cluster membership & K-means & O($N$) & \\
 & PCA & O($N$) & \\
 & kernel PCA & O($N^3$) & O($N$) \\
Photometric redshifts & NN & O($N$) & O(log$N$) \\
 & all NN & O($N^2$) & O($N$) \\
Describing a photo-z PDF & KDE & O($N^2$) & O($N$) \\
Cross-matching multi-wavelength data & nPCF$^a$ & O($N^n$) & O($N^{{\mathrm{log}}N}$) \\
Clustering of background objects & nPCF & O($N^n$) & O($N^{{\mathrm{log}}N}$) \\
\noalign{\smallskip}\hline\noalign{\smallskip}
\end{tabular}
$^a$ nPCF = n-point correlation function
\end{table}

\section{Concluding Questions}
\label{sec:questions}

The algorithms presented have excellent potential for improving astronomical analysis. Nevertheless, there are questions one can ask. We ask them here in the spirit of discussion, and to emphasize the counterpoint that the astronomer provides to the statistician (and no doubt vice-versa in other papers in this conference).

\begin{itemize}
\item{Will statistical inference (i.e., Bayesian) methods turn out to be more useful for most problems than the prediction-oriented methods presented here?}
\item{Are the approximations introduced in some of the algorithms to enable the speedups (e.g., the kernel methods), unacceptably large?}
\item{Will the algorithms be rendered insufficiently useful because of errors on the inputs?}
\item{Are the algorithms limited when the dataset does not fit in memory (either too big, or portions are run in parallel)?}
\item{Will most astronomical data analyses still contain stages that cannot be practically addressed by these algorithms, and that also scale worse than $N$log$N$, thus overwhelming even a CANFAR-like parallel computing system?}
\item{Will there be data of high {\it intrinsic} dimension, that cannot easily be dimension-reduced, thus causing curse-of-dimensionality-type problems that may hamper these algorithms?}
\item{Will novel supercomputing hardware, such as GPGPUs, that enable extremely fast brute-force approaches to problems such as nearest neighbors, prove more practical?}
\item{If the software is licensed, rather than free and open source, will it be practical to deploy it on a distributed computing system for astronomical use?}
\item{Will astronomers require the sophistication of the more advanced algorithms, or will the simple ones that scale remain `good enough', because the improvements brought by new data still account for most of the new science return?}
\end{itemize}

There are arguments one can make that the answer to all of these is ``no'', and, indeed, some are made in the manuscript, because the authors are cognizant of these questions. But, as always, if we knew all the answers, it wouldn't be research.

%
%

\end{document}